\documentclass[11pt]{article}
\usepackage[preprint]{acl}

\usepackage{times}
\usepackage{latexsym}
\usepackage[T1]{fontenc}
\usepackage[utf8]{inputenc}
\usepackage{microtype}
\IfFileExists{inconsolata.sty}{\usepackage{inconsolata}}{}
\usepackage{graphicx}
\usepackage{amsmath,amssymb}
\usepackage{bm}
\usepackage{booktabs}
\usepackage{tabularx}
\usepackage{array}
\usepackage{enumitem}
\usepackage{listings}

\defcitealias{aera2014}{AERA et al., 2014}

\title{The Computational Assessment Stack: An Architecture for AI-Native Measurement}

\author{Alina A. von Davier \\
  Duolingo}

\begin{document}
\maketitle

\begin{abstract}
Test development is shifting toward automation and agentic artificial intelligence (AI), in which autonomous systems draft items at scale, assemble test forms, administer adaptive assessments, score complex responses, and monitor test quality. As autonomous systems take on more of this work, the architecture underlying test development must change to preserve assessment quality under the new paradigm.
This paper proposes the \emph{Computational Assessment Stack} (CAS), a four-layer architecture for AI-native measurement where every stage of the assessment lifecycle shares a unifying \emph{computational construct representation} as its base layer. This representation is a machine-readable, modular, and versioned knowledge graph. Three layers are built above it: \emph{computational content generation} that builds on the knowledge graph; \emph{computational psychometrics} that derives substantive structure directly from the knowledge graph; \emph{an orchestration layer} in which agentic systems operationalize the stack under human governance.
The paper's core contribution is architectural. Building on foundational psychometric frameworks, the CAS provides a persistent representation that unifies the four layers. Unlike the descriptive prose traditionally used to define constructs in test development, the proposed architecture renders measurement assumptions inspectable and their consequences empirically testable.
\end{abstract}


\section{Introduction: From Capabilities to Architecture}

Educational measurement has entered an era in which artificial intelligence (AI) with human governance can reliably conduct many of the test development processes. AI systems are used to generate test items, score complex constructed responses, deliver adaptive assessments, and synthesize diagnostic feedback at scale. Even though each capability has been rigorously studied \citep{hao2024,vondavier2024,vondavier2026}, they have not as of yet been part of a unified architecture for AI-native assessments. What reliably links the target construct to content generation, content to psychometric calibration, calibration to score inference, and inference to the ultimate claims reported about a test taker? This progression has been presented as the chain of inferences in an argument-based framework for test validity \citep{kane1992,kane2006}. This paper investigates what \emph{sustains} and preserves that inferential chain when core lifecycle processes become automated.

In current operational practice, humans and prose-based documentation are the connective tissue that machines can act upon. Construct frameworks, test specifications, alignment studies, and calibration models are each written as descriptive prose, typically kept in separate documents, and maintained on their own, in formats machines cannot act on. Each artifact is relevant to its own stage of development, but there are not necessarily links to artifacts across stages. Humans make the connections by interpreting each document and links. As descriptive prose, these artifacts cannot easily be queried by a machine, validated systematically, version-controlled, or enforced computationally (for example, to constrain AI content generators). As AI capabilities are introduced, often unsystematically, a shared representation becomes necessary to keep the validity argument intact. 

An AI-native measurement system is one designed from the ground up for AI. It assumes that AI participates in every stage of the assessment lifecycle, autonomously or not. This includes everything from generating content to delivering tests, scoring responses, and monitoring quality. AI applications are not retrofitted onto a workflow built for human-only operation.

This paper proposes an architecture, the \emph{Computational Assessment Stack} (CAS), in which every stage of the measurement lifecycle is represented computationally and designed to work together. The CAS has four layers: a \emph{computational construct representation} at its base; \emph{computational content generation}; \emph{computational psychometrics} (delivery, estimation, and scoring); and an \emph{orchestration layer} in which agentic AI coordinates the system under human governance. The CAS is governed by a single principle: \textbf{every consequential measurement decision should have an explicit computational representation before it is automated.}

The term \emph{computational} requires precise definition here. A construct specification is \emph{recorded} when written down in a document, and made \emph{explicit} when its claims are stated rather than implied. A construct becomes \emph{computational} when an assessment system can evaluate, execute, and enforce it programmatically. The system can query construct coverage (e.g., identifying subskills lacking items), validate content against blueprints, drive targeted item generation, and maintain version control over dependent validity claims. Unlike continuous embeddings or vector spaces, this representation is symbolic (composed of discrete, typed statements programmatically evaluated as database queries). Put simply: documented constructs can be read, explicit constructs can be understood, and computational constructs can be executed. In short, the computational construct representation serves as a machine-readable test blueprint.

The CAS responds to a broader shift in how assessments are built and run. Automated item production is already established in operational programs \citep[e.g., the Duolingo English Test's item factory;][]{vondavier2024}. The more consequential change, however, is the rise of agentic AI systems that act with increasing autonomy across the entire assessment lifecycle \citep{vondavier2026}. It is this shift that makes a shared, machine-actionable construct representation necessary. Delegating workflow autonomy requires explicit, machine-readable construct representations, task models, and blueprints that agents can programmatically query, satisfy, and be held accountable to.

Accountability, in turn, requires a provenance record indicating what agents actually did. When agents generate and route content, \emph{provenance} becomes central to validity. Provenance is a machine-readable record that captures authorship, version, grounding evidence, and empirical status (expert-asserted, data-confirmed, or data-contested) and provides a chain of custody for measurement claims.

Due to the stochastic nature of AI-based content generation, we cannot fully explain why a model produced a particular item or item component. We can, however, log everything around the generation, such as the inputs and model version used, the item that was produced, the checks it passed or failed, and the decision to admit or reject it. That log is what makes each item's path into the pool traceable. The computational construct representation makes this traceability possible, serving simultaneously as the AI agent contract, operational substrate, and audit record.

\setcounter{subsection}{1}
\subsection{The Computational Assessment Stack}

\begin{figure*}[t]
  \centering
  \includegraphics[width=\textwidth]{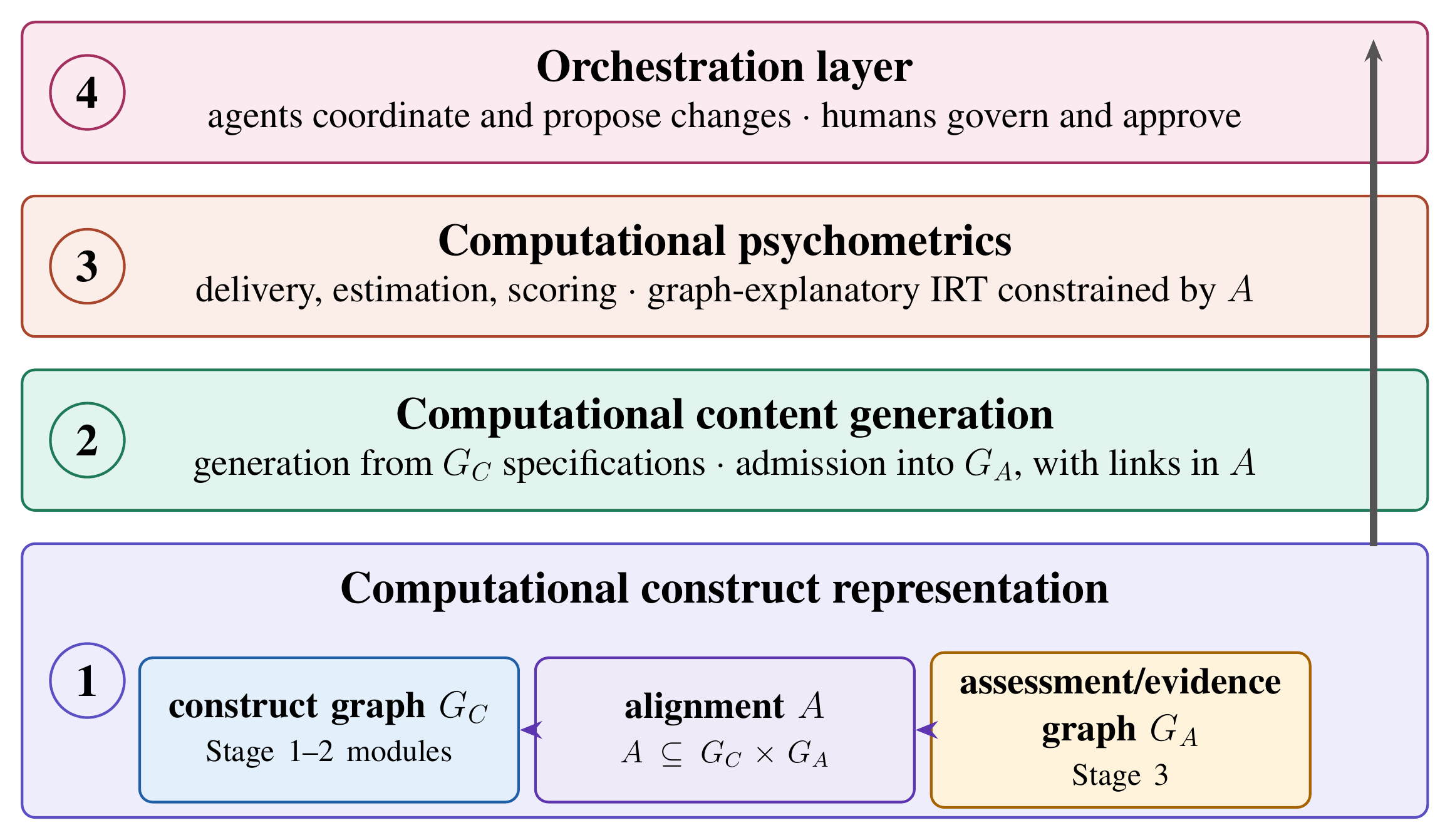}
  \caption{The Computational Assessment Stack. The computational construct representation is the connective tissue of the stack. Layer 1 comprises the construct graph $G_C$, the assessment/evidence graph $G_A$, and the alignment $A$ between them. The single arrow denotes support, not sequence. Every layer reads from, and writes evidence back to, the base. Figure~\ref{fig:governance} unpacks these flows.}
  \label{fig:cas}
\end{figure*}

The CAS's claim is that in an AI-native system, all layers should share one substrate, the construct representation at the base. This integration is the contribution of this paper, not any single layer. Figure~\ref{fig:cas} shows the architecture. The stack rests on Layer 1, the computational construct representation, which holds three objects: the construct graph $G_C$ (what is to be measured), the assessment/evidence graph $G_A$ (what is administered, observed, and inferred), and the alignment $A$ between them. The three layers above (content generation, computational psychometrics, and orchestration) each read from and write to this shared base. The single arrow in the figure denotes support, not sequence. These objects are defined formally in Section~\ref{sec:layer1}. We name the layer after the construct, its organizing object. The layer includes the construct graph $G_C$ (what the layer exists to represent), the alignment $A$ (which ties the assessment to it), and the assessment graph $G_A$ (which records how the construct is instantiated). Together the three form one versioned representation.

By sharing a unifying substrate, the computational construct representation, the generation layer draws specifications from this representation and returns items with validated links to the construct representation, the psychometric layer derives its substantive structure from the same construct representation, and the orchestration layer coordinates the other layers to ensure every reported claim throughout the CAS remains fully traceable to its foundational evidence captured in the computational construct representation.

Following the CAS's lineage (Section~\ref{sec:lineage}), we examine the foundational, computational construct representation layer that includes the construct graph, the assessment graph, and the alignment between them (Section~\ref{sec:layer1}), their modular design (Section~\ref{sec:modularity}), and the Q-matrix as part of that alignment (Section~\ref{sec:qmatrix}). Content generation, psychometrics, and orchestration follow (Sections~\ref{sec:generation}--\ref{sec:orchestration}). Section~\ref{sec:uncertainty} formalizes representation uncertainty, and Section~\ref{sec:casestudy} walks through Common European Framework of Reference (\citealp{coe2020}; CEFR) B1 Reading descriptors end to end. Finally, Section~\ref{sec:fairness} discusses fairness, Section~\ref{sec:future} sets out the validation agenda and open problems, and Section~\ref{sec:conclusion} concludes.

\section{What Already Exists: The Lineage of Each Layer}\label{sec:lineage}

A proposal for an ``AI-native architecture'' invites the question of whether the field has already built formal, machine-actionable structure into assessment. It has, repeatedly and successfully. But each prior line of work formalizes one layer, or one link between layers, and none provides a persistent representation shared across the whole lifecycle. This section reviews that work, both to credit it and to locate what the stack adds.

\subsection{Design and validity frameworks: ECD, PADI, e-ECD, and Kane's validity approach}

Evidence-centered design \citep[ECD;][]{mislevy2003} and Kane's validity framework \citeyearpar{kane1992,kane2006} provide the logical foundation for connecting claims, evidence, and validation warrants. While projects like PADI \citep[Principled Assessment Designs for Inquiry;][]{mislevy2006} encoded ECD design patterns as structured, machine-readable objects, these frameworks lack a persistent operational object, that is, a machine-readable construct representation that a running system can version, query, and enforce.

\subsection{Machine-actionable skill structures: knowledge space theory and learning maps}

Knowledge space theory \citep[KST;][]{doignon1985,falmagne2011} and learning maps \citep{kingston2016} established that formal, machine-actionable skill structures can operate at scale in programs like ALEKS and Dynamic Learning Maps assessment. While these systems prove feasibility of a formal skill structure for operational tests, they lack generality and modularity. Their underlying structures are single-purpose or bespoke, treat alternative decompositions as mutually exclusive, do not account for representation uncertainty, and were not designed to constrain AI generation. Recent work extends this line to the language domain. \citet{ribeiroflucht2026} encode CEFR A1--A2 English grammar as a typed graph of forms and functions, validate its mappings with expert annotation, and use it to guide LLM-based tutoring. Such resources supply the content of a construct representation but not its governance, meaning they are not versioned and not linked to a measurement model. Moreover, they treat their consensus graph as a single decomposition.

\subsection{Diagnostic structures: Q-matrices, attribute hierarchies, and diagnostic classification models}

The Q-matrix \citep{tatsuoka1983} bridges construct and model by specifying which skills each item requires, while attribute hierarchies \citep{leighton2004} and hierarchical diagnostic classification models \citep[DCMs;][]{templin2014} add prerequisite relations that are included in a construct graph. Rather than introducing a new statistical model, Section~\ref{sec:qmatrix} proposes an architectural generalization. Classical Q-matrices and attribute hierarchies become \emph{views} of a richer construct representation in a graph rather than hand-written tables.

\subsection{Constrained generation: cognitive design systems and automatic item generation}

Embretson's \citeyearpar{embretson1998} cognitive design systems and the early work on automatic item generation \citep[AIG;][]{bejar2002,gierl2013} pioneered item generation constrained by formal models, where cognitive templates generate families of items whose psychometric properties are predictable from design features. The CAS's generation layer (Section~\ref{sec:generation}) inherits this lineage but expands its scope. While a template constrains a single item family, the construct representation constrains every family and remains attached through calibration and reporting. Consequently, graph-constrained generation can target new subskills without new templates and enforce global rules like blueprint coverage. Furthermore, the stack strictly separates item generation from item pool admission, the governed step where items enter operational use with validated construct links.

\subsection{Feature-explained parameters: explanatory IRT}

Explanatory item response theory, from Fischer's \citeyearpar{fischer1973} linear logistic test model (LLTM) through the general framework of \citet{deboeck2004}, models item parameters as functions of item properties rather than estimating them freely. Section~\ref{sec:psychometrics}'s proposal, in which item parameters are functions of an item's position in the construct graph, sits squarely in this tradition, as does recent work that models item difficulty from item content \citep{netik2026}. The extension is representational, where the LLTM explains difficulty via hand-engineered features, the graph-explanatory model explains item parameters through a structured representation of the construct itself.

\subsection{Computational psychometrics}

Computational psychometrics \citep*{vondavier2021} blends data-driven computer science methods with theory-driven psychometric modeling across digitally mediated learning and assessment systems \citep[see also][]{vondavieryan2026}. Recent applications include LLM-based scoring, automated coding of communication data, and machine-learning parameter calibration (\citealp{sharpnack2025}, \citealp{sharpnack2026}; \citealp{hao2026}), including scalable Bayesian explanatory IRT engines designed for large banks of sparsely linked, automatically generated items \citep{nydick2026}. We use the term computational psychometrics both for the discipline and, narrowly, for the CAS's third layer, where its models reside. The stack as a whole can be read as computational psychometrics made architectural.

\subsection{What is missing}

The paper's novelty can be defined by five increasingly demanding capabilities: (i) specifying relationships among assessment components; (ii) storing and acting on formal representations in an operational system; (iii) versioning those representations and tracking provenance; (iv) carrying a single representation through generation, calibration, inference, and reporting; and (v) treating the representation itself as uncertain and evaluating competing versions. Comparing existing frameworks reveals structural asymmetry---ECD and Kane reach (i), PADI reaches (ii) at design time, ALEKS reaches (iii) for a closed structure, DLM sits between (ii) and (iii), and the L2 Network \citep{ribeiroflucht2026} reaches (ii) operationally and (iii) only partially, through its released endorsement data. To our knowledge, none combines lifecycle persistence, versioned provenance, modular revision, and representation uncertainty. The CAS claims this complete combination of a persistent, machine-actionable construct representation defined by modularity and provenance across the entire assessment lifecycle, as shown in Figure~\ref{fig:cas}.

\section{Layer 1 Foundations: What Is a Computational Construct Representation?}\label{sec:layer1}

This section defines the foundational object of the CAS, the \emph{Computational Construct Representation}.

\subsection{Definition}

We define a computational construct representation as a formal object consisting of:

\begin{enumerate}[leftmargin=*]
  \item \textbf{Entities.} Constructs, subskills, facets, text types, proficiency levels, task features, response features, scoring rules, evidence types.
  \item \textbf{Attributes.} Properties of entities, such as level, domain, language, or a distributional parameter.
  \item \textbf{Relations.} Typed, directed edges between entities, such as \emph{requires}, \emph{prerequisite-of}, \emph{aligned-to}, \emph{elicits}, \emph{evidences}. A typed relationship is an edge that includes a label saying what type of connection it is.
  \item \textbf{Constraints.} Formal conditions on the graph, such as the condition that every item must be aligned to at least one descriptor. For example, every B1 item (in a CEFR sense) must use a text type permitted at B1.
  \item \textbf{Provenance.} For each entity, relation, or constraint, a record of who or what introduced it, when, and on what basis.
  \item \textbf{Versions.} The representation changes over time, under versioning (downstream inferences reference the version that justified them).
\end{enumerate}

The first three characteristics (entities, attributes, relations) describe the \emph{content} of the representation, and most of the prior structures reviewed in Section~\ref{sec:lineage} possess them. A Q-matrix has entities and relations, a learning map has all three. The last three (constraints, provenance, versions) describe the \emph{governance} of that content, and they are what distinguish a computational construct representation from both a static specification document and its formal predecessors. The six properties characterize the representation as a whole, and the two graphs and the alignment introduced in Section~\ref{sec:alignment} all carry them.

A useful way to understand the construct graph is to think of it as a machine-readable test blueprint. Where a table of specifications records the intended test structure for human readers, the graph captures those same commitments, along with their provenance and versions, in a form the system can query and enforce. Every statement in the graph can be read as a subject--relation--object triple. In the machine learning literature model cards \citep{mitchell2019} represent a similar concept. The construct representation plays a similar documentary role for the construct, and the system itself can enforce what the record says.

A fair question is whether a graph can capture everything a careful prose specification conveys, let alone everything experts know. It cannot. Any representation is a simplification. But a prose specification is a simplification too. The difference is that the graph's simplifications are explicit, inspectable, and revisable, and what the representation omits can be named, debated, and added, module by module. Some constructs may be harder to represent as a graph altogether, as we discuss later in the paper.

\subsection{Psychometric consequences}

Each group of properties has a direct psychometric consequence. Entities, attributes, and relations make the representation \emph{queryable}, and coverage and gap analyses become computations rather than manual compilations. Constraints make it \emph{enforceable}, and a misaligned item is caught at screening instead of discovered retrospectively. Provenance makes it \emph{auditable}, and the basis for any alignment judgment or decomposition can be examined and challenged. Versions make it \emph{traceable}, and therefore, every claim links to the representation and model versions that supported it, which matters in high-stakes programs where frameworks, specifications, and models change over time. The case study presented later in this paper illustrates each capability in turn.

None of the consequences represents validity evidence. Queryability tells us where the gaps are, not whether the categories are right. Enforcement tells us an item conforms to the blueprint, not that the blueprint is sound. Auditability and traceability tell us \emph{which} claims were made under \emph{which} assumptions, not that the assumptions were warranted. The properties create the objects on which validation can operate and the validation remains to be done.

\subsection{Two graphs, one relation}\label{sec:alignment}

Throughout the rest of the paper we distinguish two graphs. The \textbf{construct graph}, $G_C$, contains the entities, attributes, and relations that constitute the construct, that is, what is to be measured. The \textbf{assessment/evidence graph}, $G_A$, contains the entities that constitute the assessment (items, task features, response features, scoring rules, psychometric parameters, observations, evidence, inferences), what is administered, observed, and inferred. \textbf{Alignment} is a relation between the two graphs, representing a set of links, each pairing an element of $G_C$ with the elements of $G_A$ that are claimed to instantiate it. We write $A \subseteq G_C \times G_A$. The relation is not intrinsic to either graph. It is a set of claims about how the assessment instantiates the construct. A construct graph can be specified before any item exists, and an assessment graph accumulates content as items are written, administered, and calibrated. Alignment is the mapping between them.

\paragraph{What the alignment contains.} The links in $A$ are typed, and they are of two kinds. \emph{Construct-alignment} links connect an item to the parts of the construct it is claimed to measure (an item \emph{targets} a subskill, or is \emph{aligned to} a framework descriptor). \emph{Operational-conformance} links connect an item to the specifications it is built to satisfy, that is, an item instantiates a task template, or uses a text type permitted at its target level. The classical Q-matrix is a slice of the first kind. Write $s_1, \ldots, s_K$ for the subskill nodes of the active Stage 1 module. Restricting $A$ to subskill--item pairs gives $q_{ik}=1$ exactly when $(s_k, i) \in A$ under the \emph{targets} relation (Section~\ref{sec:qmatrix}). Relations that stay within one graph are not alignment, meaning that a prerequisite between two subskills belongs to $G_C$, and the response features of an item belong to $G_A$.

\paragraph{Asserted and empirical alignment.} Every link in $A$ begins as a claim, asserted by an expert panel or proposed by an AI generator and accepted in review. Response data can then confirm or contest it. We therefore distinguish the \emph{asserted} alignment, which the system enforces at screening and which can fix the structure of a psychometric model (the hard constraints of Section~\ref{sec:graphconstrained}), from its \emph{empirical status}, which records whether the data support each link (the provenance labels of Section~\ref{sec:hybrid} and the soft constraints of Section~\ref{sec:graphconstrained}). In these terms, empirical Q-matrix validation \citep{delatorre2008,chiu2013,delatorre2016} is empirical validation of one slice of $A$.

\paragraph{Alignment as a measurement problem.} Many alignment questions can be \emph{formulated} as graph comparison problems. Alignment questions arise at two levels. At the level of a single link, the question is whether a claimed correspondence holds: whether an item meets with the specification it was written to, or whether its content corresponds semantically to the descriptor it is aligned to. At the level of a pool or a form, the question is how well $G_A$ as a whole instantiates $G_C$, as in blueprint coverage. For the second level, given a target construct graph $G_C$, an assessment graph $G_A$, and the alignment $A$, one can define a family of similarity functions combining node coverage, relational agreement, prerequisite-path agreement, and distributional agreement over content categories, with different weights. Any such function is a \emph{candidate metric}. Before interpretation, its reliability, sensitivity to relevant graph changes, and invariance across populations need to be established.

\subsection{Design decisions}

Building a computational construct or assessment representation requires several decisions.

\paragraph{Implementation technology.} Several standard technologies can host the computational representation. One is an RDF (Resource Description Framework) knowledge graph \citep{w3crdf}, a web-standard format in which every statement is a subject--relation--object triple. RDF comes with SHACL (Shapes Constraint Language; \citealp{w3cshacl}), a language for automated checks on graph structure that acts as automated blueprint enforcement, and SKOS (Simple Knowledge Organization System; \citealp{w3cskos}), a vocabulary for recording correspondences to external frameworks. Alternatives are a property graph in a graph database, or a relational database schema with integrity constraints. The machine-level encoding of the case study for implementers is given in the Appendix. The paper's argument rests on the general claim that the construct becomes a computational object inside the assessment system (stored, queried, checked, and versioned by the same infrastructure that manages items, parameters, and scores) rather than a document consulted alongside it, hence it is independent of any one of these technologies.

\paragraph{Authoring procedure.} How the graph is authored (how its content is elicited and encoded) is itself part of the validity argument. Candidate procedures include consensus-based expert decomposition, independent-then-aggregated expert decomposition, and LLM-extracted decomposition (with or without expert validation). Each has different properties with respect to reliability, coverage, and defensibility, and the authoring procedure used should be documented in the representation's provenance.

\paragraph{Grain size.} A finer-grained representation is more informative but harder to link to standard statistical models. While a binary Q-matrix is coarse and easy to model, it discards substantive information. This is the familiar question of attribute grain size from the diagnostic-modeling literature, extended to the whole representation. This paper takes no position on the optimal grain size. The claim here is that the modular architecture proposed in the next section makes the trade-off explicit, inspectable, and revisable.

\section{Modularity: Designing Against Ossification}\label{sec:modularity}

\subsection{The ossification problem}

Once a construct is formalized as a graph, three risks emerge: provisional claims harden into structural commitments, alternative decompositions become invisible, and low-level revisions trigger cascading invalidations across items, calibrations, and reports. Relying solely on human governance (``humans approve changes'') does not solve how to make revision affordable. Without explicit architectural design, a computational representation risks becoming \emph{less} flexible than the prose it replaces. This is a neglected risk that partly explains why bespoke operational graphs remain difficult to maintain and compare.

\subsection{A modular architecture}\label{sec:modular}

\begin{figure*}[t]
  \centering
  \includegraphics[width=\textwidth]{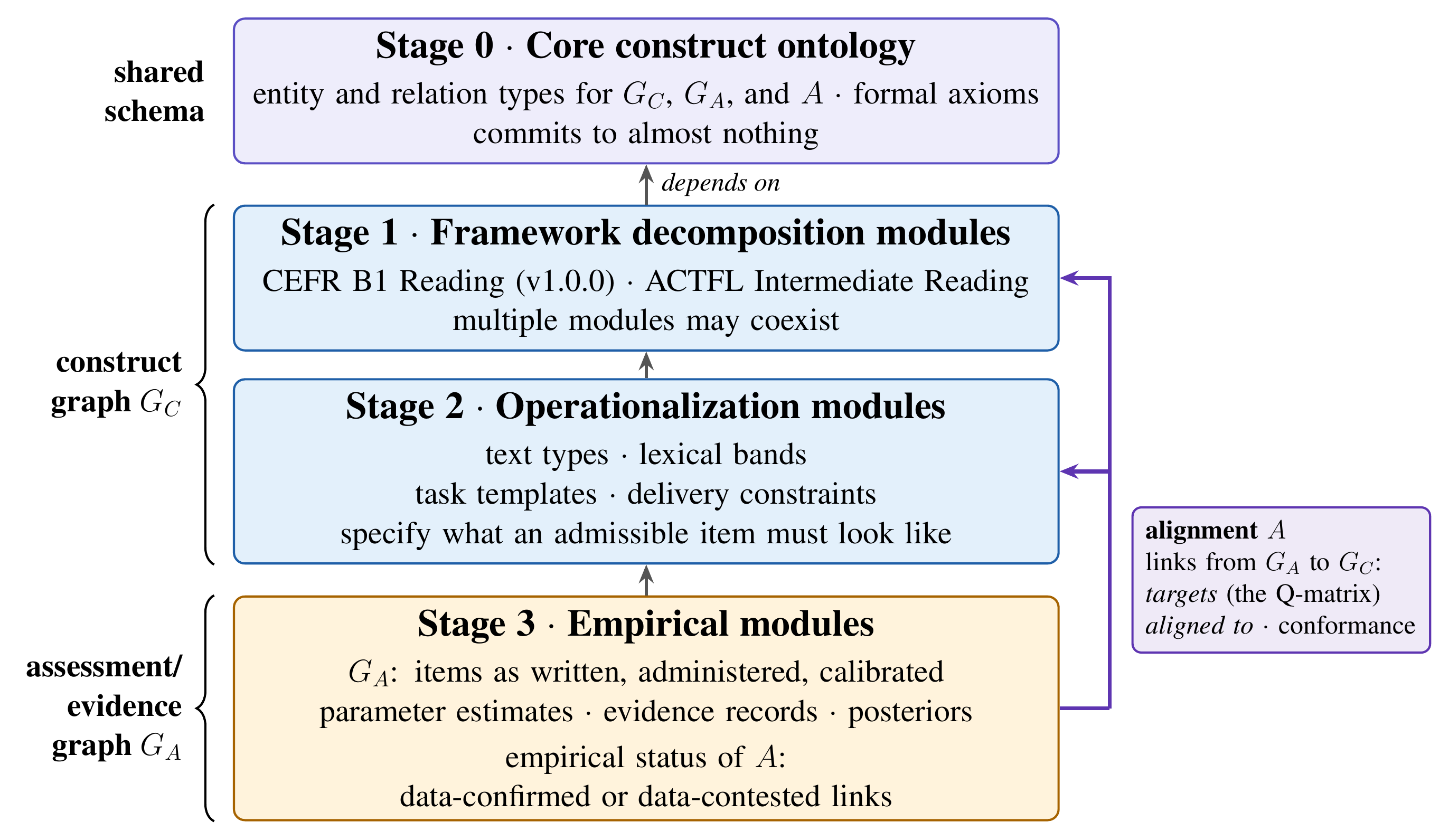}
  \caption{The modular architecture. Stage 0 is the shared schema; Stages 1 and 2 form the construct graph $G_C$; Stage 3 holds the assessment/evidence graph $G_A$ and the empirical status of the alignment $A$, whose links run from $G_A$ to $G_C$. Arrows point to what a stage depends on; stability is greatest at the top and volatility greatest at the bottom; revisions propagate downward and are bounded by stage. Multiple modules may coexist at every stage, each Stage 1 module $m$ carrying its own alignment $A_m$.}
  \label{fig:modular}
\end{figure*}

We propose a modular architecture for $G_C$, $G_A$, and $A$, organized into four stages, each independently replaceable, governed by a design principle that dependencies flow downward. Higher stages commit to less. Lower stages are replaceable without invalidating the stages above them. The architecture maps onto Cronbach and Meehl's \citeyearpar{cronbach1955} construct--observable distinction and Kane's argument-based validity, where each stage transition represents an inference with a corresponding warrant. Figure~\ref{fig:modular} summarizes the four stages.

\begin{itemize}[leftmargin=*]
  \item \textbf{Stage 0: Core construct ontology.} The schema defining the kinds of entities and relations that $G_C$, $G_A$, and $A$ may contain, not any specific content. For example, defining allowable entities, relation types (including the types of alignment link), and formal axioms (e.g., that every construct has at least one facet). This stage commits to almost nothing, remaining stable across framework revisions.
  \item \textbf{Stage 1: Framework decomposition modules.} It includes decompositions of a core construct into framework-specific subskills (e.g., CEFR B1 Reading, or Intermediate Reading in the Proficiency Guidelines of the American Council on the Teaching of Foreign Languages, ACTFL, \citeyear{actfl2024}). Multiple modules may coexist, each versioned and validated separately. Stage 1 modules belong to $G_C$.
  \item \textbf{Stage 2: Operationalization modules.} It includes context-specific task templates, text types, lexical bands, and delivery constraints that depend on Stage 1 modules without being entailed by them. Stage 2 modules also belong to $G_C$ because they specify what an admissible item must look like, while the items themselves belong to $G_A$.
  \item \textbf{Stage 3: Empirical modules.} It includes $G_A$ (the items as written, administered, and calibrated, with their parameter estimates, evidence records, and posteriors) together with the empirical status of each link in $A$, such as an item-to-subskill link that the data confirm or contest. Stage 3 instantiates the upper stages without ever modifying them.
\end{itemize}

Formally, let $G^{(l)}$ denote the set of modules at stage $l$, each module carrying its own version. The construct graph is the union of the active Stage 1 and Stage 2 modules, $G_C = G^{(1)} \cup G^{(2)}$, written in the schema $G^{(0)}$; Stage 3 holds $G_A$ and the empirical status of $A$. Dependencies flow $G_A \rightarrow A \rightarrow G_C$, and within $G_C$, $G^{(2)} \rightarrow G^{(1)} \rightarrow G^{(0)}$, that is an item depends on the links that align it, and those links depend on the construct elements they point to, never the reverse. Three properties follow:

\begin{itemize}[leftmargin=*]
  \item \textbf{Local revision.} Replacing a Stage 2 module leaves Stage 0 and Stage 1 in place; only the links in $A$ that point to the replaced module, and the items and estimates in $G_A$ that depend on those links, are flagged for re-validation. The cost of a revision is bounded by the stage in which it occurs.
  \item \textbf{Composability.} Multiple modules can coexist at each stage; the graph is a lattice, not a tree, meaning that a subskill or operationalization may serve several parents, and alternatives may sit side by side rather than one decomposition excluding all others. Each coexisting Stage 1 module $m$ carries its own alignment $A_m$, which is what allows inferences to be compared, or combined, across modules (Section~\ref{sec:uncertainty}).
  \item \textbf{Traceability.} Every claim is traceable to a specific configuration of module versions in $G_C$, the links in $A$ it relied on, and the objects in $G_A$ that supplied its evidence.
\end{itemize}

\section{The Q-Matrix as a View of the Alignment}\label{sec:qmatrix}

The Q-matrix is part of the alignment $A$ (Section~\ref{sec:alignment}), being the slice of $A$ that pairs items with subskills. The graph representation therefore generalizes the \emph{representational scope} of the Q-matrix rather than replacing it. It embeds item--skill incidence within a richer set of typed relations, and a Q-matrix can then be derived as a view of that representation and supplied to a compatible psychometric model. A Q-matrix, a prerequisite hierarchy, and an item-to-domain mapping can coexist in one representation. However, this generalization is representational, not statistical.

The classical Q-matrix \citep{tatsuoka1983} is a binary incidence structure $Q = (q_{ik})$, with $q_{ik} = 1$ if item $i$ requires skill $k$ and 0 otherwise. It bridges the construct and the psychometric model. The representation embeds this incidence matrix within a family of typed relations, which differ in where they sit:

\begin{itemize}[leftmargin=*]
  \item \textbf{Item--subskill incidence (in $A$).} $q_{ik} \in \{0, 1\}$: whether item $i$ requires subskill $s_k$; these are the \emph{targets} links between $G_A$ and the Stage 1 modules of $G_C$.
  \item \textbf{Item--domain incidence (in $A$).} $d_{id} \in \{0, 1\}$: whether item $i$ belongs to content domain $d$.
  \item \textbf{Skill prerequisite structure (within $G_C$).} $p_{jk} \in \{0, 1\}$: whether subskill $s_j$ is a prerequisite for subskill $s_k$ --- the relation modeled by the attribute hierarchy method \citep{leighton2004} and hierarchical DCMs \citep{templin2014}. This structure will need both feasibility checks and structural constraints to keep the dependencies computable. A graph can better handle dependencies.
  \item \textbf{Item--evidence-type incidence (within $G_A$).} $e_{ie} \in \{0, 1\}$: whether item $i$ elicits evidence type $e$ (selected response, constructed response, process).
\end{itemize}

The classical Q-matrix is the special case in which $P$, $E$, and $D$ are trivial, and the attribute hierarchy is the special case in which $E$ and $D$ are trivial. The generalization is that the Q-matrix is one relation in a family of typed relations, the family is the substantive specification, and each relation carries provenance and a version. Where a relation sits matters. For example, $Q$ and $D$ are claims that cross from the assessment to the construct and inherit the measurement status of $A$, $P$ is a claim about the construct alone, and $E$ describes the assessment alone.

Because graph relations could serve different statistical roles, the mapping from graph to model must be explicitly declared. A graph-derived Q-matrix inherits standard identifiability challenges, such as the completeness and non-redundancy conditions established in the DCM literature \citep{gu2021}, and must be validated against expert-derived and empirically derived alternatives \citep{delatorre2008,chiu2013}. Such validation asks whether the \emph{targets} links in $A$ are supported by response data. Ultimately, the graph's contribution is not to replace these established methods, but to provide a richer, relational, provenance-bearing specification against which they can be rigorously compared.

\subsection{Hybrid construction: expert elicitation refined by data}\label{sec:hybrid}

The computational construct representation should not rest on expert judgment alone. Experts propose the initial structure grounded in theory and curriculum, keeping the validity argument non-circular, while assessment response data can test the edges in $G_C$ and the links in $A$ via established methods like empirical Q-matrix validation \citep{delatorre2016} and Bayesian network structure learning \citep{heckerman1995}. Because data-driven structures reflect population and teaching sequences as much as constructs (if skill X is always taught before skill Y, the data will suggest a prerequisite whether or not one truly exists), provenance tracking (with labels such as expert-asserted, data-confirmed, and data-contested) ensures challenged edges and links trigger expert review rather than silent revision, strengthening the validity argument. A recent CEFR grammar graph illustrates an alternative approach. \citet{ribeiroflucht2026} retain a form--function mapping in their released core only if at least two experts endorse it, publishing the full endorsement data separately. In the CAS, such minority mappings would remain in the graph as data-contested edges or as a coexisting module rather than being filtered out.

\section{Layer 2: Computational Content Generation}\label{sec:generation}

With the construct graph in hand, item generation becomes a constrained generation problem in the lineage of cognitive design systems and AIG \citep{embretson1998,gierl2013}, with the shared construct representation replacing the per-family template as the constraining object. Generation proceeds by querying the graph for the target descriptor, subskills, and permissible task features, then prompting a generator (an LLM, an agentic workflow, or a template-based script) with the resulting subgraph as context.

The pipeline is: Specify $\rightarrow$ Generate $\rightarrow$ Structural screening (checks of form and metadata) $\rightarrow$ Semantic evaluation (checks of meaning) $\rightarrow$ Human review $\rightarrow$ Pilot $\rightarrow$ Calibration $\rightarrow$ Admission to the item pool. Screening and semantic evaluation run automatically, before any human review. This ordering is the design proposal, and operational programs already implement parts of it.

We emphasize the distinction between \emph{generation} and \emph{admission}. Generation is unconstrained production, meaning that a generator can produce any string, and prompting alone does not reliably yield proficiency-aligned content \citep{benedetto2025}. Admission is governed measurement, that is, the step by which an item enters the operational pool with explicit, validated links to the construct. The construct graph constrains admission, not generation. Structural screening runs the graph's admission constraints mechanically on every candidate as soon as it is generated. Admission proper is the final step, after review, piloting, and calibration. The consequence is that generation can be fast, exploratory, and cheap, while admission is slow, constrained, and documented, and the two can be run by different models, different agents, and different governance.

The Item Factory \citep{vondavier2024} provides an operational workflow for this separation of generation and admission. Subject-matter experts design tasks, AI generates items at scale, humans review for quality and fairness, differential item functioning (DIF) analyses monitor for bias, and monitoring continues after launch. The construct graph leaves this workflow intact. It makes the links between generation, review, and calibration explicit and machine-traceable. When an admitted item later exhibits DIF or misfit, the graph makes it possible to ask \emph{which construct claims} the item was admitted under, and whether other items admitted under the same claims share the problem. Over a graph, that question is a simple query, while over a traditional document trail, it is a complicated and time-consuming process.

\section{Layer 3: Computational Psychometrics in the CAS}\label{sec:psychometrics}

The construct graph specifies the substantive structure, which is what the psychometric work should be about. This section proposes graph-constrained models, graph-explanatory models, and a model comparison that makes the framework falsifiable.

\subsection{Graph-constrained psychometric models}\label{sec:graphconstrained}

Suppose the construct graph identifies $K$ latent dimensions, one per subskill node, where dimension $k$ is the ability measured by subskill $s_k$ (Section~\ref{sec:alignment}). For person $p$, let $\bm{\theta}_p=(\theta_{p1},\ldots,\theta_{pK})^\top$ denote the vector of latent abilities, and for item $i$, let $\mathbf q_i=(q_{i1},\ldots,q_{iK})^\top$, where $q_{ik}\in\{0,1\}$ indicates whether the graph specifies that the item measures dimension $k$.

Suppose an item is intended to measure Reading and Vocabulary but not Listening.

Its graph-derived Q-matrix row might be
\[
\mathbf q_i = (1,1,0).
\]
The discrimination vector is
\[
\mathbf a_i = (a_{i1}, a_{i2}, a_{i3}).
\]
Under a hard graph constraint, the model specifies
\[
\mathbf a_i = (a_{i1}, a_{i2}, 0).
\]
Only the first two components are estimated, while the third is fixed at zero. This means the item response probability cannot depend directly on the person's Listening ability, conditional on the other modeled abilities.

The vector $\mathbf q_i$ is therefore a graph-derived specification of the item's permitted dimensions, that is, its Q-matrix row, the item's slice of the alignment $A$ (Section~\ref{sec:qmatrix}).

Under a \textbf{hard graph constraint}, the item's Q-matrix row decides which abilities its response probability may depend on. Wherever the row has a zero, the item's discrimination on that dimension is fixed at zero:
\[
q_{ik}=0 \implies a_{ik}=0.
\]
Wherever it has a one, the discrimination is estimated from response data, subject to the model's identification constraints. The data determine how large the permitted discriminations are, and the item difficulty $b_i$ remains freely estimated unless it is explicitly modeled from graph-derived properties (Section~\ref{sec:explanatory}). A multidimensional two-parameter logistic (2PL) model is then
\[
P(X_{pi}=1\mid\bm{\theta}_p)
=\operatorname{logit}^{-1}
\left(\mathbf a_i^\top\bm{\theta}_p-b_i\right),
\]
where $X_{pi}$ is person $p$'s response to item $i$, $b_i$ is the item difficulty parameter, and $\mathbf a_i^\top\bm{\theta}_p$ is the item's discrimination-weighted combination of the person's abilities on the permitted dimensions. Thus, the graph specifies which dimensions may contribute to an item response, while the data determine the strength of those contributions and the item difficulty.

The model requires standard identification constraints. For example, the latent means may be fixed at zero, the latent variances at one, and the latent covariance matrix $\bm{\Sigma}$ estimated subject to these constraints. Alternatively, suitable anchor items can be used to identify the latent scales. Discriminations are constrained to be positive on permitted dimensions. These constraints must be sufficient to identify the model. The existence of $K$ graph dimensions does not guarantee that all $K$ dimensions are empirically distinguishable in a particular item pool. Dimensional identifiability is therefore an empirical and design requirement, not an assumption guaranteed by the representation.

The resulting model can be compared with less constrained alternatives to assess whether the proposed structure adequately represents the response data.

A \textbf{soft graph constraint} treats the graph's dimensional assignments as prior information rather than absolute restrictions. Discriminations on permitted dimensions receive priors consistent with the chosen identification constraints, while discriminations on excluded dimensions are allowed to be nonzero but receive zero-centered priors. For example,
\[
a_{ik}\sim
\begin{cases}
p_{\mathrm{allowed}}(a_{ik}), & q_{ik}=1,\\
\mathcal N(0,\tau^2), & q_{ik}=0,
\end{cases}
\]
where $p_{\mathrm{allowed}}$ is a prior supported on positive discriminations, and $\tau$ controls the degree of shrinkage toward zero on excluded dimensions. The prior scale must be appropriate to the identification of the latent variables. Under this formulation, the graph expresses an initial hypothesis about the item's dimensional structure, but response data can provide evidence of additional relationships.

The distinction is substantive. Hard constraints test a construct representation that makes categorical claims about which dimensions an item measures. Soft constraints allow those claims to be challenged by empirical evidence. In either case, the graph provides a prespecified structural hypothesis, but it does not establish that the hypothesis is correct.

\subsection{The graph as an explanatory representation}\label{sec:explanatory}

Graph-constrained IRT specifies which dimensions an item may measure but leaves its permitted discrimination parameters and difficulty to be estimated from response data. A stronger formulation uses the construct graph to explain the item parameters themselves. In this formulation, the graph is not merely a constraint on the psychometric model, but it is the representation from which item parameters are predicted.

Let $G_i$ denote the graph-based representation of item $i$. It includes the relevant construct nodes and relationships activated by the item, such as the depth and specificity of the skills involved, their prerequisite relationships, and the item's position relative to proficiency levels. A graph-based explanatory model maps this representation to two outputs:
\[
f_\phi(G_i)=
\left(\mathbf a_\phi(G_i),b_\phi(G_i)\right),
\]
where $f_\phi$ is a function with parameters $\phi$, $\mathbf a_\phi(G_i)$ is a $K$-dimensional vector of predicted discrimination parameters, and $b_\phi(G_i)$ is a scalar predicted difficulty parameter. The function $f_\phi$ is learned from calibrated items and its outputs are the item parameters used in the response model.

The response model is therefore
\begin{multline*}
P(X_{pi}=1\mid\bm{\theta}_p,G_i)\\
=\operatorname{logit}^{-1}
\left[
\mathbf a_\phi(G_i)^\top\bm{\theta}_p-b_\phi(G_i)
\right].
\end{multline*}
Under a hard graph constraint, the predicted discrimination vector must respect the graph's permitted dimensions, so $q_{ik}=0$ implies $a_{\phi,k}(G_i)=0$. Under a soft constraint, the mapping can instead incorporate the graph as prior information, allowing excluded dimensions to receive small, nonzero discriminations. The same distinction between structural restrictions and prior beliefs therefore applies to the explanatory formulation.

The baseline model should be a hierarchical explanatory IRT model using prespecified graph-derived features. For example, item difficulty might be modeled as
\begin{align*}
b_i={}&\beta_0+
\beta_1\operatorname{Depth}(G_i)\\
&+\beta_2\operatorname{PrerequisiteComplexity}(G_i)\\
&+\beta_3\operatorname{SkillDistance}(G_i)+
\varepsilon_i,
\end{align*}
where the graph-derived predictors are defined before model fitting and $\varepsilon_i$ captures residual item-level variation. The three predictors proposed here are standard node-level graph measures. $\operatorname{Depth}(G_i)$ is the level of the item's targeted subskills in the prerequisite hierarchy; $\operatorname{PrerequisiteComplexity}(G_i)$ is the number of distinct subskills required by the targeted subskills \citep{falmagne2011}; and $\operatorname{SkillDistance}(G_i)$ is the graph distance (shortest-path length) between the item's targeted subskills, capturing how dispersed the item's demands are across the construct graph. These are illustrative operationalizations computed from $G_C$ and the item's \emph{targets} links in $A$; the exact definitions are themselves design choices to be recorded with the model's provenance. Other graph-measures may be considered and tested. Discrimination parameters can be modeled analogously, subject to appropriate positivity, dimensionality, and identification constraints. This baseline is interpretable and can be estimated using standard psychometric methods.

This formulation extends the explanatory IRT tradition. The Linear Logistic Test Model \citep[LLTM;][]{fischer1973} is a special case in which item difficulty is expressed as a linear function of coded cognitive operations. \citet{deboeck2004} provide a broader framework for explanatory item response models. The proposed extension is to use a structured, versioned representation of the construct graph as the explanatory basis for item parameters, rather than relying exclusively on manually engineered item-feature vectors.

Three distinct prediction tasks should be distinguished:

\begin{enumerate}[leftmargin=*]
  \item \textbf{Parameter prediction for calibrated items:} estimating item parameters from explanatory information and evaluating them against parameters obtained from response data.
  \item \textbf{Parameter recovery from a reference calibration:} assessing whether the explanatory model can recover parameters established by an independent or reference calibration.
  \item \textbf{Prediction for genuinely new items:} predicting item parameters before any response data are available for those items.
\end{enumerate}

The third task is the most demanding deployment scenario. Strong performance on held-out items from familiar templates does not establish that the model can generalize to genuinely novel items or item families.

The proposed architecture is therefore
\[
G_i
\to
(\mathbf a_i,b_i)=f_\phi(G_i)
\to
\text{IRT model}
\to
X_{pi}.
\]
The graph explains the item parameters, and the IRT model retains the measurement structure that relates those parameters to person abilities and observed responses. This differs from directly predicting item difficulty or response accuracy from the graph without an explicit measurement model. In the latter case, the graph may support prediction, but its role in explaining the measurement process is less clearly specified.

For newly generated items, graph-predicted parameters should initially be treated as provisional estimates or as the basis for informative priors, not as established calibration results. As response data accumulate, the parameters can be updated through a hierarchical model, consistent with item-family calibration approaches \citep{glas2003,deboeck2004}. This approach is also consistent with the proposal that predicted parameters for new items serve as provisional priors that are subsequently updated with calibration evidence \citep{sharpnack2026}, and with operational calibration engines built for this setting, in which feature-informed priors for automatically generated items are updated as sparse response data accumulate \citep{nydick2026}.

\subsection{A falsifying comparison: four models}\label{sec:fourmodels}

The explanatory claim becomes testable through a comparison of four models. The comparison asks whether the construct graph provides sufficient information to explain item parameters, and whether it adds value beyond conventional item features.

\begin{itemize}[leftmargin=*]
  \item \textbf{M0 --- Conventional IRT:} Item discrimination and difficulty parameters are estimated freely from response data. This model provides a reference for fit and predictive performance when item parameters are calibrated directly, rather than predicted from explanatory variables.
  \item \textbf{M1 --- Feature-explanatory IRT:} Item parameters are modeled as functions of conventional item features, $(\mathbf a_i,b_i)=f_\phi(\mathbf z_i)$, where $\mathbf z_i$ is a vector of conventional item features (linguistic, content, format).
  \item \textbf{M2 --- Features-plus-graph explanatory IRT:} Item parameters are modeled using both conventional features and graph representations, $(\mathbf a_i,b_i)=f_\phi(\mathbf z_i,G_i)$.
  \item \textbf{M3 --- Graph-only explanatory IRT:} Item parameters are modeled from the graph representation alone, $(\mathbf a_i,b_i)=f_\phi(G_i)$.
\end{itemize}

Here $f_\phi$ denotes an explanatory mapping in the sense of Section~\ref{sec:explanatory}, with parameters $\phi$ estimated separately under each model. The models differ only in what information the mapping is allowed to use, and $\mathbf a_i$ and $b_i$ denote its outputs, the item parameters used in the response model.

The principal comparison is M3 against M2, with M1 establishing the predictive value of conventional explanatory features and M0 providing a reference based on freely estimated item parameters. If M3 performs comparably to M2 on appropriate out-of-sample evaluations, this would support the claim that the graph captures much of the information in conventional item features that is relevant to the modeled item parameters. It would not, by itself, establish that the graph captures every construct-relevant property of an item.

If M3 underperforms M2, the difference is diagnostically useful. Conventional features may contain construct-relevant information missing from the graph, suggesting that the representation needs to be enriched. Alternatively, their apparent predictive advantage may reflect construct-irrelevant properties, suggesting a need for tighter feature screening. The gap may also reflect limitations of the mapping function $f_\phi$, rather than deficiencies in the graph itself. These possibilities require a separate investigation.

For newly generated items, predicted parameters should be treated as uncertain and updated as response evidence accumulates. The graph can record each item's parameter provenance and calibration status. For example, it can record \emph{graph-predicted}, \emph{provisionally calibrated}, or \emph{empirically calibrated} item parameters along with the model and graph versions used to obtain the estimates. These labels describe the evidence supporting the parameters, not guaranteed levels of accuracy. 

\section{Layer 4: Agentic Orchestration and Governance}\label{sec:orchestration}

\begin{figure*}[t]
  \centering
  \includegraphics[width=\textwidth]{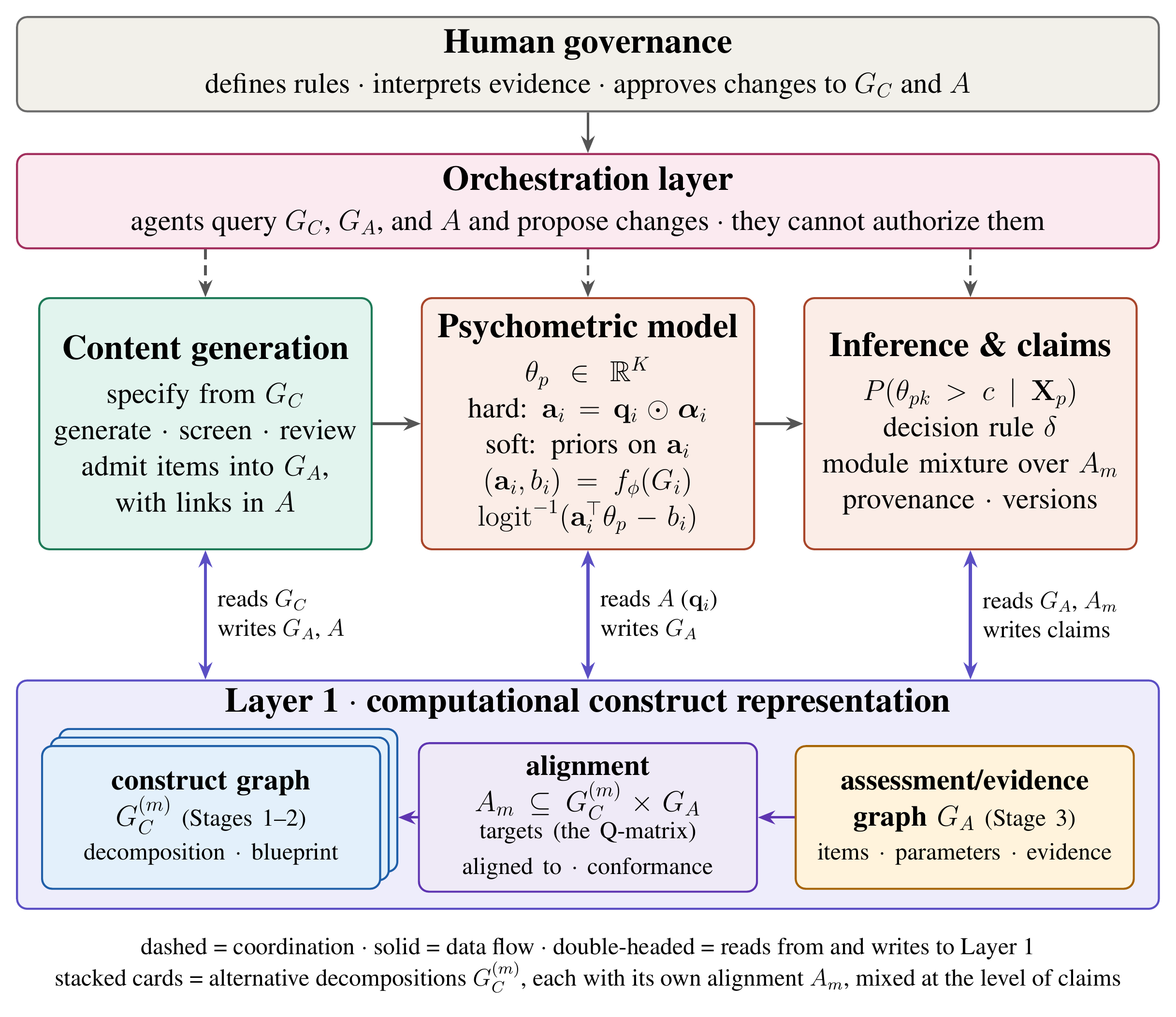}
  \caption{The Computational Assessment Stack with human governance and data flows. Content generation, the psychometric model, and inference each read from and write to Layer 1, which holds $G_C$ (stacked: alternative Stage 1 modules $G_C^{(m)}$), the alignment $A_m$ of each module, and $G_A$.}
  \label{fig:governance}
\end{figure*}

The orchestration layer coordinates assessment workflows, such as content generation, item recalibration, and quality review, using agentic AI under human governance. The principles underlying the orchestration are:

\begin{enumerate}[leftmargin=*]
  \item \textbf{Construct-Authority Principle:} Agents may query the representation and propose changes to the construct graph $G_C$ and the alignment $A$, but cannot autonomously authorize them. Human approval remains mandatory for changes to what is measured and to what counts as measuring it (Figure~\ref{fig:governance}).
  \item \textbf{Modular Governance:} Agents propose versioned module updates across stages (e.g., alternative decompositions or operationalizations), which require appropriate human approval before becoming operational:

  module version $v$ $\rightarrow$ (agents query and propose) $\rightarrow$ module version $v+1$ $\rightarrow$ (human approval) $\rightarrow$ operational.
  \item \textbf{Centralized Provenance:} Every reported claim (e.g., mastery classifications) is fully reconstructible, linking active representation versions, measurement models, item administrations, and decision rules.
\end{enumerate}

\section{Construct-Representation Uncertainty}\label{sec:uncertainty}

The modular architecture reveals a new source of uncertainty. This is the uncertainty about the construct representation itself. Different expert panels may propose different decompositions of a competency, different frameworks may identify different subskills, and the literature may support more than one interpretation. This uncertainty occurs at two levels.

\paragraph{Edge-level uncertainty.} A single relationship in the graph (a prerequisite between two subskills, say) may or may not be supported by the available evidence. This level is handled by machinery already introduced. The provenance labels of Section~\ref{sec:hybrid} (expert-asserted, data-confirmed, data-contested) record each edge's current status, and the evidence trail records how that status was reached and revised. A contested edge triggers expert review rather than silent deletion.

\paragraph{Module-level uncertainty.} The harder case arises when whole decompositions disagree. Two Stage 1 modules may differ in the number of subskills, their definitions, or their organization, and both may be \emph{admissible}, that is grounded in theory, endorsed by experts, and consistent with the data so far. When admissible modules coexist, the system has three responses, in increasing order of ambition.

First, \emph{report per module}. The default is that every claim already carries, in its provenance, the module version and alignment $A_m$ under which it was computed (the Centralized-Provenance principle of Section~\ref{sec:orchestration}), so claims from different modules can coexist without being merged.

Second, \emph{report what is robust}. Sensitivity analysis recomputes consequential claims (proficiency classifications, decisions about individuals) under each admissible module and asks whether they change. Hence, the recommendation is to report conclusions only at the level of detail that the available representations support consistently. If the modules agree on an overall proficiency classification but differ in how they define particular subskills, the overall classification is defensible while subskill-level claims require qualification.

Third, \emph{combine the modules' predictions}. Combination is legitimate only for a claim that means the same thing under every module being combined. This means that there must be an explicitly defined decision rule that can be applied to each module's posterior distributions, together with a documented mapping from each module's outputs to the common claim. An overall proficiency classification may qualify, if each module supports the same classification and the thresholds have a defensible shared interpretation. A claim about a particular subskill generally does not qualify, when the modules define their subskills differently. Whether the meanings correspond is an expert judgment, not a computation, and where no correspondence can be established the modules are evaluated separately. When the condition holds, the combined probability of a claim $C_p$ about person $p$ is a weighted average over the candidate modules $M_m$:
\[
P(C_p\mid X_p)=
\sum_m P(C_p\mid X_p,M_m)\,w_m,
\]
where $P(C_p\mid X_p,M_m)$ is the probability of the claim computed under module $m$'s alignment and measurement model, and the weights $w_m$ are nonnegative and sum to one. Following predictive stacking \citep*{yao2018}, the weights are estimated by optimizing the combination's out-of-sample predictive performance. The right analogy is weighting forecasters by their track records, not voting on which forecaster is right. Bayesian model probabilities address which candidate model generated the data. Stacking weights measure how much each module contributes to prediction (and a module can receive a large weight while remaining substantively wrong about the construct).

\paragraph{What remains subjective.} The formula does not remove expert judgment. The weights can only redistribute belief among the modules that panels actually wrote, and no mixture can represent a decomposition nobody proposed. Whether $C_p$ means the same thing under $M_1$ and $M_2$ is decided by expert review, and combining without that warrant produces a number with no interpretation. And the weights do not indicate the probability that a module is correct. Evaluating representation uncertainty therefore requires complementary analyses, where experts establish which representations are defensible, predictive evaluations measure how well each performs, and sensitivity analysis determines which conclusions depend on the choice.

\paragraph{Cross-framework comparison.} A more difficult case arises when assessments use different frameworks, such as the CEFR and ACTFL. If the frameworks define different constructs or make different claims, their outputs cannot simply be combined using the mixture above. Comparison first requires a crosswalk that specifies how claims in one framework correspond to claims in the other. One possible approach is to estimate a mapping between their Stage 1 construct graphs, allowing correspondences to vary in strength rather than assuming exact equivalence. Such a mapping would require substantive validation and, where available, empirical evidence from a linking sample. Developing and validating cross-framework mappings remains an open problem.

\paragraph{Invariance of the representation.} A parallel question arises across populations and language versions. Measurement invariance asks whether a psychometric model behaves equivalently across groups; \textbf{construct-representation invariance} asks whether the representation itself does. If construct graphs for the reading inference exist in English and in Spanish, $G_C(\text{English})$ and $G_C(\text{Spanish})$, are they structurally equivalent? Semantically? Measurement-equivalent? These are distinct questions with direct fairness implications (Section~\ref{sec:fairness}), and the sensitivity analysis above complements traditional measurement-invariance testing by asking whether the representation, both its dimensions and relationships, remains sufficiently consistent for the intended comparisons.

\section{An End-to-End Case Study: A Fraction of CEFR B1 Reading}\label{sec:casestudy}

This section walks through a hypothetical specification example following a small slice of the CEFR B1 Reading framework through the full lifecycle, from official descriptors and computational decomposition to item generation, screening, calibration, and audit logging. All numerical values are illustrative. The CEFR is a reference framework that the Council of Europe describes as supporting transparency, coherence, and comparability across curricula, assessments, and languages. Our representation preserves the distinction between its official descriptors and our computational interpretation. Readers interested in the machine-level encoding will find it in the Appendix. Figure~\ref{fig:casestudy} shows the full example as a graph. The walk-through instantiates the stages of Figure~\ref{fig:modular}. The schema (Stage 0) is taken as given. Steps 2 and 3 build the Stage 1 and Stage 2 modules that make up $G_C$. Steps 5--7 populate Stage 3 --- the items and evidence of $G_A$, and the empirical status of their links in $A$. This content is presented in specification tables, mirroring traditional tables of specifications, since every statement in the graph can be read as a row of the form \emph{subject --- relation --- object} --- the machine-readable analogue of a row in a table of specifications. The example's layering (subskills, permissible text types, task templates) is the assessment analogue of the form--function--task layering used in computational-linguistics grammar graphs \citep{ribeiroflucht2026}. While this resource stops at the domain model, the walk-through example continues through calibration, evidence, and a reported claim.

\begin{figure*}[t]
  \centering
  \includegraphics[width=\textwidth]{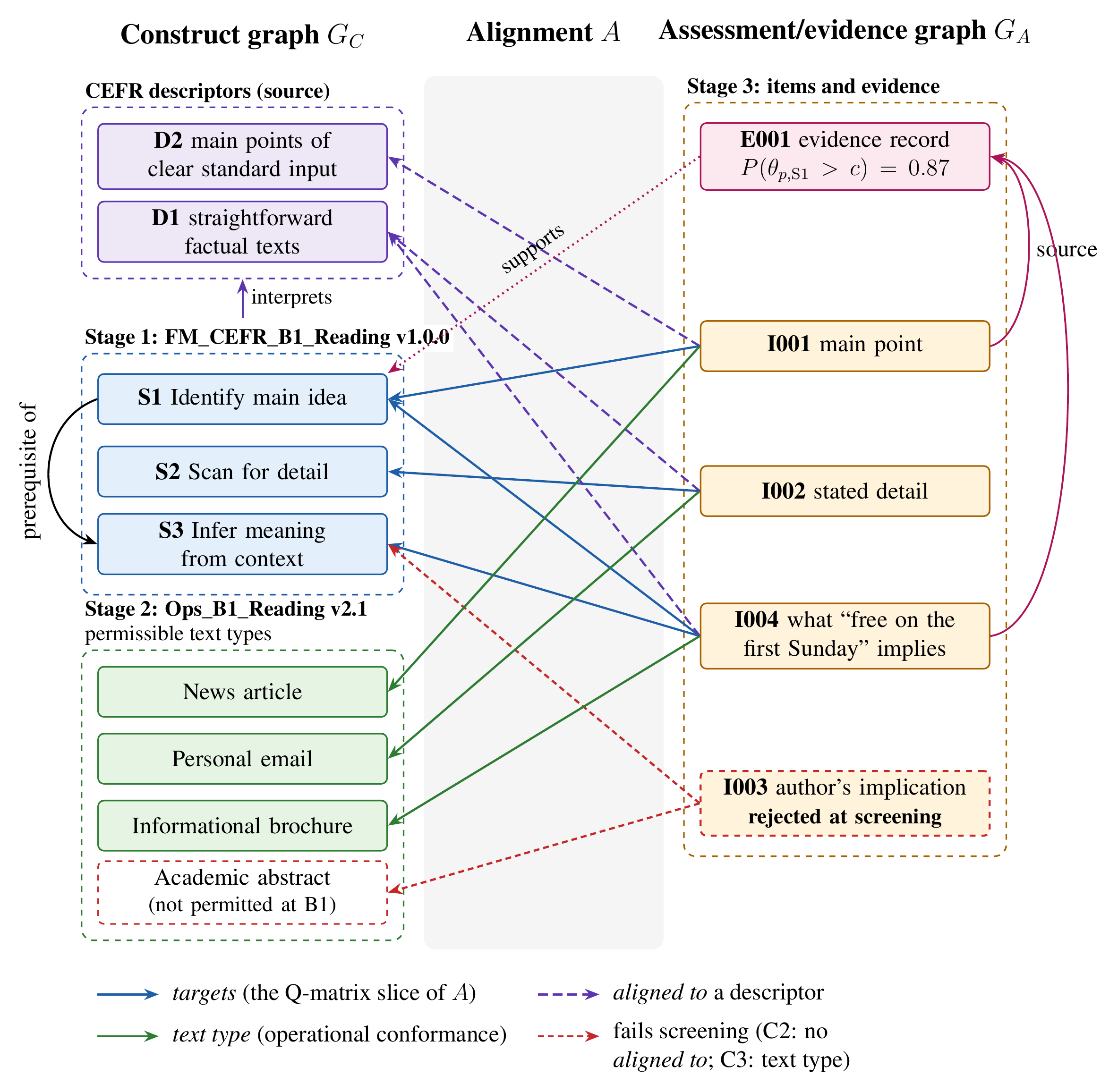}
  \caption{The CEFR B1 Reading case study, arranged as the two graphs and the alignment between them. Left: the construct graph $G_C$, made up of the Stage 1 decomposition module and the Stage 2 operationalization module, with the CEFR descriptors interpreted by the Stage 1 module. Right: the assessment/evidence graph $G_A$ (Stage 3), with the four candidate items and the evidence record E001. Middle: the alignment $A$, whose links are of two kinds --- construct alignment (\emph{targets}, the item--subskill links that form the Q-matrix, and \emph{aligned to}) and operational conformance (\emph{text type}). I004 targets both S1 and S3, as the panel judged (Section~\ref{sec:step6}). Item I003 is rejected at screening because it has no \emph{aligned-to} link (C2) and its text type is not permitted at B1 (C3); see Table~\ref{tab:screening}.}
  \label{fig:casestudy}
\end{figure*}


\subsection{Step 1: Source descriptors}\label{sec:step1}

We begin with two illustrative B1 Reading descriptors, cited by reference rather than reproduced:

\begin{itemize}[leftmargin=*]
  \item \textbf{D1:} Reading straightforward factual texts on subjects related to the learner's field and interests with a satisfactory level of comprehension.
  \item \textbf{D2:} Understanding the main points of clear standard input on familiar matters regularly encountered in work, school, or leisure.
\end{itemize}

These are paraphrases for illustration. A working system would reference the official descriptor identifiers in the Companion Volume \citep{coe2020}.

\subsection{Step 2: Decomposition as a Stage 1 framework module}\label{sec:step2}

An expert panel interprets the two descriptors as jointly requiring three subskills, S1, S2, S3 (the nodes $s_1, s_2, s_3$ of Section~\ref{sec:alignment}, so $K=3$), with one prerequisite relation. The interpretation is the panel's, not the Council of Europe's, and the module records that explicitly (Table~\ref{tab:stage1}). In an operational module, the panel would also draw on published resources that specify linguistic realizations by level, the Council of Europe's Reference Level Descriptions and, for English, the English Profile program \citep{ribeiroflucht2026}, and the module's provenance would record those sources alongside the panel.

\begin{table*}[t]
  \centering
  \small
  \begin{tabularx}{\textwidth}{@{}l>{\raggedright\arraybackslash}p{0.27\textwidth}l X@{}}
    \toprule
    \textbf{Statement} & \textbf{Subject} & \textbf{Relation} & \textbf{Object} \\
    \midrule
    1 & Module FM\_CEFR\_B1\_Reading, v1.0.0 & interprets & Descriptors D1, D2 \\
    2 & Module FM\_CEFR\_B1\_Reading & proposes subskills & S1: Identify main idea; S2: Scan for detail; S3: Infer meaning from context \\
    3 & S1: Identify main idea & prerequisite of & S3: Infer meaning from context \\
    4 & Module FM\_CEFR\_B1\_Reading & provenance & Expert panel, 2025-09-01; consensus procedure \\
    \bottomrule
  \end{tabularx}
  \caption{The Stage 1 framework decomposition module}
  \label{tab:stage1}
\end{table*}

The module is a \emph{proposal} about decomposition. A second panel might read the same descriptors differently (organizing them around reading processes, say, or splitting S3 into lexical and pragmatic inference) and that reading would enter the graph as a coexisting module with its own identity. (Versions record a single module's own history and coexisting modules record genuinely different readings of the construct.) The choice between coexisting modules is a substantive research question (Sections~\ref{sec:uncertainty} and~\ref{sec:future}), not a technical one.

\subsection{Step 3: Operationalization as a Stage 2 module}\label{sec:step3}

A Stage 2 module operationalizes the decomposition for a particular delivery context. It functions as the machine-readable portion of the test blueprint, that includes permissible stimulus types, task templates, and the admission constraints (Table~\ref{tab:stage2}).

\begin{table}[t]
  \centering
  \small
  \begin{tabularx}{\columnwidth}{@{}>{\raggedright\arraybackslash}p{0.33\columnwidth}X@{}}
    \toprule
    \textbf{Element} & \textbf{Specification} \\
    \midrule
    Permissible B1 text types & Short news article; personal email; informational brochure \\
    Not permissible at B1 & Academic abstract; technical manual \\
    Task template T1 & Four-option multiple choice; exactly one key; stimulus 150--220 words \\
    Constraint C1 & Every item has exactly one keyed option \\
    Constraint C2 & Every item is aligned to at least one official descriptor \\
    Constraint C3 & Every item's text type is permitted at the item's target level \\
    Constraint C4 & Every item targets at least one subskill in an active Stage 1 module \\
    Blueprint rule R1 & Every subskill in the active module has at least two admitted items \\
    \bottomrule
  \end{tabularx}
  \caption{The Stage 2 operationalization module}
  \label{tab:stage2}
\end{table}

Constraints C1--C4 are the admission constraints, enforced mechanically at the screening step (Section~\ref{sec:step5}). They are the kinds of rules every test specification already contains. What changes is that they are \emph{enforced}, automatically, on every candidate item, rather than checked by hand on a sample.

\subsection{Step 4: A blueprint coverage query}\label{sec:step4}

Because the blueprint and the item pool live in the same representation, coverage is a query. Suppose the current pool contains three admitted items for S1 and two for S2 (both satisfying blueprint rule R1) and none for S3. The standing query ``which subskills of the active module violate rule R1?'' returns exactly one row, S3, which is a gap. The orchestration layer flags it and issues a generation request targeted at S3, carrying with it the relevant subgraph, that includes the subskill definition, its prerequisite (S1), the permissible text types, and task template T1. In conventional practice this gap analysis is a periodic content-mapping exercise; here it runs continuously as the pool changes.

\subsection{Step 5: Generation and screening}\label{sec:step5}

Stage 3 begins here. From this step on, the example adds items and evidence to $G_A$ and alignment links to $A$, leaving $G_C$ unchanged. An item AI generator (an LLM in this example) is prompted with the S3 subskill information in the graph and produces candidate items. The pool already holds two previously drafted candidates. For I004, the panel asserts two \emph{targets} links. The item targets S3 (the subskill the generation request named) and also requires its prerequisite S1, so its asserted Q-matrix row is $(1, 0, 1)$ over (S1, S2, S3). Item I003 is deliberately flawed, with an academic-abstract stimulus (a text type not permitted at B1) and no alignment link to any official descriptor. Nothing prevents a generator from producing such an item, which is the point of this step. Every candidate is checked mechanically against the constraints C1--C4 before any human sees it (Table~\ref{tab:screening}).

\begin{table*}[t]
  \centering
  \small
  \begin{tabularx}{\textwidth}{@{}l X l >{\raggedright\arraybackslash}p{0.27\textwidth} >{\raggedright\arraybackslash}p{0.13\textwidth}@{}}
    \toprule
    \textbf{Item} & \textbf{Stimulus and task} & \textbf{Target} & \textbf{Screening result} & \textbf{Disposition} \\
    \midrule
    I001 & News article (180 words); MC: best statement of the main point & S1 & Passes C1--C4 (aligned to D2) & To human review \\
    I002 & Personal email (160 words); MC: locate a stated detail & S2 & Passes C1--C4 (aligned to D1) & To human review \\
    I003 & \emph{Academic abstract} (190 words); MC: what the author implies about zoning & S3 & \textbf{Fails C2} (no descriptor link); \textbf{fails C3} (text type not permitted at B1) & \textbf{Rejected at screening} \\
    I004 & Informational brochure (170 words); MC: what ``free of charge on the first Sunday'' implies for a Saturday visitor & S3, S1 & Passes C1--C4 (aligned to D1) & To human review \\
    \bottomrule
  \end{tabularx}
  \caption{Candidate items and screening against admission constraints C1--C4}
  \label{tab:screening}
\end{table*}

I003 is rejected automatically, with a machine-generated rejection record naming the violated constraints and the module versions in force, before item review, before piloting, before any examinee sees it. In a document-based workflow, the same flaws would be caught, if at all, by a reviewer who happened to compare the item against the right page of the specification.

Note that the checks verify \emph{form}, not \emph{substance}. They confirm that I004 has exactly one keyed option, not that the key is correct, and that it is linked to a descriptor, not that the link is appropriate, etc. Also note that an item can pass every structural check and still be a bad item. The correctness of the key is confirmed in a separate review (human or a trained reviewer AI agent). Whether the item measures what it claims is a question for pilot data and calibration.

\subsection{Step 6: Piloting and calibration}\label{sec:step6}

I001, I002, and I004 pass human review and are piloted. Suppose they are piloted within a larger pool, say 40 items spanning the three subskills, with roughly 2,000 responses. Three items alone could not identify a three-dimensional model, and the quantities below are hypothetical calibration outputs, invented for illustration rather than estimated from response data. The model is the graph-constrained multidimensional 2PL of Section~\ref{sec:graphconstrained}, with three correlated dimensions (S1, S2, S3), compensatory, under the hard constraint $q_{ik}=0 \Rightarrow a_{ik}=0$, with the identification constraints given there. ``Loads on'' in Table~\ref{tab:calibration} names the dimensions on which the asserted Q-matrix permits a nonzero discrimination. In a real calibration the data would estimate the values. I004 loads on S1 as well as S3, following its asserted row $(1,0,1)$ from Step 5 (the module records S1 as a prerequisite of S3, and the panel judged the item to require both). Whether a prerequisite should induce a loading is itself a modeling decision, one the representation makes explicit.

\begin{table*}[t]
  \centering
  \small
  \begin{tabularx}{\textwidth}{@{}l l >{\raggedright\arraybackslash}p{0.25\textwidth} l X@{}}
    \toprule
    \textbf{Item} & \textbf{Loads on} & \textbf{Discrimination $a$ (SE)} & \textbf{Difficulty $b$ (SE)} & \textbf{Fit note} \\
    \midrule
    I001 & S1 & 1.10 (0.12) & $-0.35$ (0.09) & Consistent with intended loading (hypothetical) \\
    I002 & S2 & 0.95 (0.11) & $-0.10$ (0.10) & Consistent with intended loading (hypothetical) \\
    I004 & S1, S3 & 0.60 (0.14) on S1; 1.25 (0.16) on S3 & $+0.70$ (0.12) & Consistent with intended loading (hypothetical) \\
    \bottomrule
  \end{tabularx}
  \caption{Graph-derived loadings and hypothetical calibration outputs (not estimated from data)}
  \label{tab:calibration}
\end{table*}

Had I004's responses instead loaded primarily on S2, the graph--model disagreement diagnostics of Section~\ref{sec:fourmodels} would flag it. The item, the decomposition, the model, or the data would be candidates for fault, and the representation localizes the investigation.

\subsection{Step 7: Evidence, decision, and a reported claim with provenance}\label{sec:step7}

An examinee, $p$, responds to a form containing these and other items. The measurement model yields a posterior for each subskill. For S1, say, a posterior mean of 0.82 on the mastery scale with posterior standard deviation 0.14, and a posterior probability $P(\theta_{p,\mathrm{S1}}>c\mid\mathbf X_p)=0.87$ that the examinee exceeds the B1 threshold $c$ on S1. The reporting rule in force, $\delta$, classifies ``at or above B1'' on a subskill when that posterior probability is at least 0.80. The system reports the claim, and stores, with it, everything required to reconstruct it (Table~\ref{tab:provenance}).

\begin{table}[t]
  \centering
  \small
  \begin{tabularx}{\columnwidth}{@{}>{\raggedright\arraybackslash}p{0.40\columnwidth}X@{}}
    \toprule
    \textbf{Provenance element} & \textbf{Value} \\
    \midrule
    Claim & Examinee $p$ at or above B1 on S1 (Identify main idea) \\
    Posterior probability above threshold & 0.87 \\
    Posterior mean (SD) & 0.82 (0.14) \\
    95\% credible interval & [0.51, 0.97] \\
    Measurement model & Graph-constrained MIRT (2PL; Section~\ref{sec:graphconstrained}), calibration v3, 2026-02 \\
    Items contributing & I001, I004, \ldots \\
    Construct module & FM\_CEFR\_B1\_Reading v1.0.0 \\
    Operationalization module & Ops\_B1\_Reading v2.1 \\
    Decision rule & $\delta$: classify if posterior probability $\geq 0.80$ \\
    \bottomrule
  \end{tabularx}
  \caption{Hypothetical provenance record for a reported claim}
  \label{tab:provenance}
\end{table}

The claim is a decision. Note the width of the credible interval in the table and its lower end, 0.51, which is lower than any plausible mastery bar. This reflects the familiar subscore problem, that a formally represented subskill is not necessarily a reliably measured one. The provenance record makes the claim auditable, not precise, and it keeps it auditable years later.

\subsection{What the example shows}

End to end, the representation shows several things that a traditional specification document cannot provide. It computed a blueprint gap as a standing query (\ref{sec:step4}). It constrained what the AI generator was asked for and rejected a nonconforming item mechanically, with a reasoned record, before human review (\ref{sec:step5}). It bound calibration results to the decomposition that produced the loadings (\ref{sec:step6}). Last but not least, it bound a reported claim to the exact versions of the decomposition, blueprint, model, and decision rule that supported it (\ref{sec:step7}).

It did not do the psychometric work. Everything still requires review and validation.

\section{Discussion}

\subsection{Fairness considerations}\label{sec:fairness}

We distinguish five sources of bias in a graph-driven assessment system, the first two made visible by the representation layer. They are \emph{construct representation bias} (what is included or omitted), \emph{decomposition bias} (whose structure is adopted), \emph{generation bias} (whether generated content disadvantages subgroups), \emph{measurement bias} (traditional differential item functioning), and \emph{feedback bias} (discrepant diagnostic pathways driven by model error).

While formal representations make certain biases auditable in ways prose cannot, they also introduce risks at the ontology and decomposition layers. Auditing such systems connects to a growing literature on algorithmic audits \citep{raji2019,raji2020}. This indicates that fairness should be evaluated not only at the response-to-score layer \citepalias{aera2014}, but also at the construct-representation layer. Because culturally specific decompositions can embed bias before items are ever generated, the representation layer must be inspected directly, with \emph{construct-representation invariance} serving as the formal instrument for that evaluation. Such checks might also help with the design of socioculturally responsive assessments \citep{bennett2023}.

Here, we propose a framework for investigation that distinguishes three forms of equivalence. They are \emph{structural} (same nodes and relations), \emph{semantic} (same meanings), and \emph{measurement} \citep[same parameters;][]{meredith1993,millsap2011}. While traditional measurement invariance assumes a single shared representation across subgroups, \emph{computational-construct-representation invariance} extends this hierarchy to the representation itself. For example, subgroup differences in prerequisite edges constitute structural evidence rather than item-level Differential item functioning \citep[DIF;][]{holland1993}. In cross-language adaptation \citep{itc2017}, structural equivalence can be evaluated pre-calibration using graph metrics (e.g., edit distance, subgraph isomorphism). Graph structure and measurement parameters can sometimes dissociate, which is why both structural and psychometric evidence must be evaluated together.

\subsection{Validation agenda and open problems}\label{sec:future}

The CAS generates testable hypotheses (H), and it identifies problems that are not yet testable. We organize both by the object they concern (the construct graph $G_C$, the alignment $A$, and the assessment/evidence graph $G_A$) followed by the generation and governance layers that act on all three. 

\paragraph{The construct graph, $G_C$.}

\textbf{H1a, inter-expert agreement.} Can expert construct specifications be represented computationally with acceptable inter-expert agreement? Given a set of CEFR descriptors, do independent experts working with a shared schema produce decompositions whose nodes and relations agree well above chance? In the closest published precedent, expert form--function mappings for CEFR A1--A2 grammar reached moderate agreement (Krippendorff's $\alpha = 0.58$), with disagreement mostly additive rather than contradictory \citep{ribeiroflucht2026}. Note that ``experts'' could be highly trained AI agents in the future.

\textbf{H1b, empirical distinguishability.} Can competing decompositions be distinguished empirically? Agreement is not correctness. Do alternative Stage 1 modules produce measurably different downstream results in alignment scores, difficulty predictions, model fit, or reported classifications? If the results are indistinguishable, the mixture of Section~\ref{sec:uncertainty} collapses harmlessly, and if they differ, the choice of module is consequential and must be defended.

\textbf{Open problems.} a) Some constructs are harder to represent as static graphs (interactive, collaborative, dynamic constructs, and process constructs). They require representations that capture temporal and interaction structure, not merely node--edge structure. b) H1a and H1b are a beginning, but the field lacks a general framework for evaluating alternative decompositions and for validating a \emph{module} rather than a whole graph.

\paragraph{The alignment, $A$.}

\textbf{H2, alignment.} Does graph-derived alignment agree with independent expert judgments, and are misalignments detected with acceptable sensitivity and specificity? The alignment metrics discussed earlier need to be validated. 

\textbf{Open problems.} The conditions under which a graph-derived Q-matrix --- the item--subskill slice of $A$ --- is identifiable for a given DCM or MIRT model need to be established, extending \citet{gu2021} to the derived-view setting.

\paragraph{The assessment/evidence graph, $G_A$, and inference.}

\textbf{H3, psychometric structure.} Does graph-derived structure explain empirical item-response behavior? The test is the four-model design of Section~\ref{sec:fourmodels}, with performance on novel AI-generated items the most diagnostic criterion. 

\textbf{H5, inference and reporting.} Do graph-linked diagnostic claims show greater construct traceability and interpretive validity than conventional subscores, without losing psychometric precision? The graph may improve interpretability and traceability without improving reliability, since the two are distinct.

\textbf{Open problems.} The graph-to-parameter mapping $f$ of Section~\ref{sec:explanatory} is a potential source of systematic error distinct from the graph. The trade-off between interpretability, flexibility, and misspecification risk needs to be further explored. 

\paragraph{Generation and governance.}

\textbf{H4, generation.} Do graph-constrained generation systems produce items with better construct coverage and fewer substantive misalignments than unconstrained generation? A strong experiment compares unconstrained, prompt-constrained, and graph-constrained generation on expert alignment, quality, diversity, psychometric characteristics, and rejection rate at screening.

\textbf{Open problems.} What is the right division of labor between human and AI agents in construct revision, and how should agentic decisions be audited?

\section{Conclusion}\label{sec:conclusion}

The Computational Assessment Stack unifies an AI-native measurement system around a foundational, machine-readable construct representation. By integrating modular knowledge graphs with content generation, explanatory psychometrics, and agentic orchestration with human governance, the stack transforms disparate assessment components into a single, traceable architecture.

While building on established theory, from evidence-centered design to explanatory IRT, CAS's core contributions are architectural: a persistent, versioned representation across the lifecycle, a modular design preventing ossification, and a formal treatment of representation uncertainty. Ultimately, AI optimizes within the layers, but humans govern what it is.

The next frontier in assessment is not simply automating item generation or scoring, but building the computational representations that connect construct theory, task design, evidence, statistical inference, and reporting, and then validating them.

\section*{Acknowledgment}
This work is based on the invited keynote at the AI in Measurement Conference (AIME-Con) 2026. The work is a research collaboration with Anthropic's Claude Enterprise, Fable 5, with multiple interactions from June to October 2026. The author also acknowledges the use of OpenAI's GPT-6 (accessed in June 2026 and October 2026) for brainstorming and paper reviews.

Additionally, the author is grateful to Dr. Jill Burstein and Dr. Steven Nydick from Duolingo for their feedback on previous versions of this manuscript.

The author takes full responsibility for the final content, interpretations, and conclusions.

\bibliography{references}

\appendix
\onecolumn

\section*{Appendix: Machine-Level Encoding of the Case Study}

This appendix gives the RDF/Turtle encoding (Turtle is a human-readable format for writing RDF graphs; \citealp{w3cturtle}) of the Stage 1 module and the evidence record presented as tables in Sections~\ref{sec:step2} and~\ref{sec:step7}. The encoding is illustrative. It uses placeholder example.org namespaces and covers the example's objects, not a complete operational schema. Each block is a set of subject--relation--object statements; nothing here adds content beyond the tables in the body.

\medskip\noindent\textbf{The Stage 1 framework module (Section~\ref{sec:step2}):}

\begin{lstlisting}
@prefix decomp: <https://example.org/decomp#> .
@prefix cefr:   <https://example.org/cefr#> .

decomp:FM_CEFR_B1_Reading_v1
    a decomp:FrameworkModule ;
    decomp:coreConstruct cefr:Reading ;
    decomp:framework cefr:CEFR ;
    decomp:level "B1" ;
    decomp:sourceDescriptor cefr:OfficialDescriptorD1, cefr:OfficialDescriptorD2 ;
    decomp:proposes decomp:IdentifyMainIdea ;
    decomp:proposes decomp:ScanForDetail ;
    decomp:proposes decomp:InferMeaning ;
    decomp:version "1.0.0" ;
    decomp:provenance "Expert panel, 2025-09-01; consensus procedure" .

decomp:IdentifyMainIdea a decomp:SubSkill .
decomp:ScanForDetail    a decomp:SubSkill .
decomp:InferMeaning     a decomp:SubSkill .

decomp:IdentifyMainIdea decomp:prerequisiteOf decomp:InferMeaning .
\end{lstlisting}

\medskip\noindent\textbf{The screening constraints (Section~\ref{sec:step3}), expressed as SHACL shapes covering C1--C4:}

\begin{lstlisting}
decomp:B1ItemShape
    a sh:NodeShape ;
    sh:targetClass decomp:B1Item ;
    sh:property [                              # C1
        sh:path decomp:keyedOption ;
        sh:minCount 1 ; sh:maxCount 1 ;
        sh:message "Item must have exactly one keyed option." ;
    ] ;
    sh:property [                              # C2
        sh:path decomp:alignedTo ;
        sh:minCount 1 ;
        sh:message "Item must be aligned to at least one official descriptor." ;
    ] ;
    sh:property [                              # C3
        sh:path decomp:textType ;
        sh:minCount 1 ;
        sh:in ( decomp:NewsArticle decomp:PersonalEmail decomp:Brochure ) ;
        sh:message "Item must declare a text type permitted at B1." ;
    ] ;
    sh:property [                              # C4
        sh:path decomp:targets ;
        sh:minCount 1 ;
        sh:class decomp:SubSkill ;
        sh:message "Item must target at least one subskill in an active Stage 1 module." ;
    ] .
\end{lstlisting}

Run against candidate item I003 (Section~\ref{sec:step5}), this shape produces two violation reports, on C2 and C3: the mechanical rejection shown in Table~\ref{tab:screening}. Note that C3 requires a text-type value to exist (\texttt{sh:minCount 1}); without that clause, an item carrying no text type would escape the permitted-values check. The shape enforces C1 through C3 fully and the targeting half of C4. The ``active module'' half of C4 is not expressible in SHACL alone; it is enforced by one SPARQL query \citep{w3csparql} run at the same screening step:

\begin{lstlisting}
ASK {
  ?item decomp:targets ?s .
  ?m decomp:proposes ?s ;
     decomp:status "active" .
}
\end{lstlisting}

An item for which this query returns false violates C4. The division of labor is itself informative: shape constraints check an item's own structure, while queries check its relation to the current state of the graph.

\medskip\noindent\textbf{The evidence record (Section~\ref{sec:step7}):}

\begin{lstlisting}
decomp:E001
    a decomp:Evidence ;
    decomp:supports            decomp:IdentifyMainIdea ;
    decomp:posteriorMean       0.82 ;
    decomp:posteriorSD         0.14 ;
    decomp:credibleInterval    "[0.51, 0.97]" ;
    decomp:posteriorAboveCut   0.87 ;
    decomp:decisionRule        decomp:Delta_Classify_080 ;
    decomp:measurementModel    decomp:Model_MIRT_v3 ;
    decomp:source              decomp:I001, decomp:I004 ;
    decomp:constructVersion    "FM_CEFR_B1_Reading_v1.0.0" ;
    decomp:opsVersion          "Ops_B1_Reading_v2.1" .
\end{lstlisting}

\end{document}